# BRIDGE COUPLER FOR APT*

Paul T. Greninger, Henry J. Rodarte, General Atomics, San Diego, CA

*Abstract*

The Coupled Cavity Drift Tube Linac (CCDTL) used in the Accelerator for the Production of Tritium (APT) is fully described elsewhere [1]. The modules are composed of several machined and brazed segments that must account for the accumulation of dimensional tolerances in the build up of the stack. In addition, space requirements dictate that power fed to the accelerator cannot be through the accelerating cavities. As well, we would like to remove a single segment of the accelerator without removing additional segments. These requirements combined with phasing relationships of the design and space limitations have resulted in a different bridge coupling method used in the module comprising 3-gap segments. The coupling method, phasing relationships and other features that enhance the flexibility of the design will be discussed.

## 1 BRIDGE COUPLER DESIGN

The Bridge Coupler below addresses all of the above problems. A unique feature is the ability to take up tolerances, and employ the design over a wide range of particle velocities. The center cavity is bent into a U, where the distance between the legs may vary, while maintaining a total constant length. This accommodates different spacing between Accelerator Cavities (AC). The bridge coupler consists of an odd number of cavities in order to preserve the $\pi/2$ operating mode of the RF structure. Figure 1 shows the basic bridge coupler design. The concept uses $TM_{01}$ mode pillbox cavities, with axes parallel to the accelerator center-line. A unique aspect of this structure is that due to phasing requirements, the center cavity is no longer a pill box cavity, but has been replaced with a waveguide operating in a $TE_{013}$ mode. Figure 2 defines the relative phase differences, shown with arrows, for the segments being coupled. The two Coupling Cavities (CC) (see Fig. 1) are unexcited in the $\pi/2$ mode, while the center cavity is excited to a power level determined by the relative sizes of the coupling slots shown. Since the center cavity is excited, RF power can be fed directly into it through a slot coupled to a waveguide.

*Work supported under contract DE-AC04-96AL89607

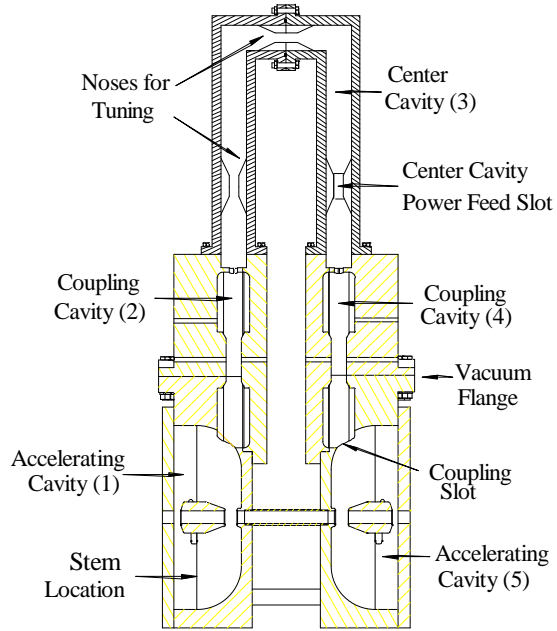

Figure 1. Drawing of Bridge Coupler Cross Section.

## 2 FIELD PATTERN

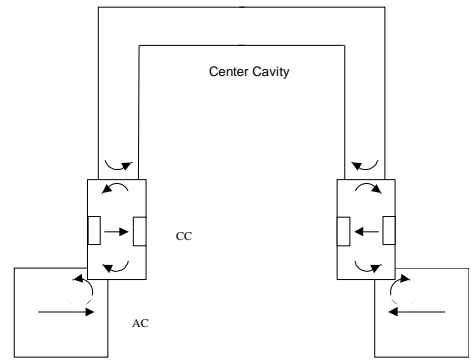

Figure 2. Field Pattern Inside Bridge Coupled Model.

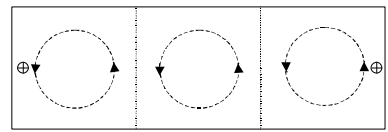

Figure 3. Field Pattern in $3/2\ \lambda_g$ Length Waveguide.

The fields from the AC couple in the same direction as the CC. The fields from the center cavity couple in opposite directions so there is zero energy in the CC's. The waveguide in Fig. 3 has its ends folded down. This gives the required 180° phase change at the guide.

## 3 MAIN FEATURES

- Coupling cavities and the center cavity have noses used for tuning.
- The coupling cavities are split along a plane parallel to the accelerator center-line. A vacuum flange is seated here. Once disassembled a single segment of the accelerator can be removed without removing additional segments.
- The center cavity is a waveguide bent into a U shape. The distance between the ACs can vary, while keeping the waveguide length and the resonant frequency unchanged.
- The center cavity noses nearest the coupling slots enhance the fields locally. This increases the coupling between the center cavity and CCs.
- The center cavity resonates in the $TE_{013}$ mode in order to provide the correct relative $180^0$ RF phasing of the accelerating cavities. The tuning noses are located at electric field maxima.
- The relative sizes of the center cavity tuning noses near the coupling slots can be adjusted to mix in a lower order mode ($TE_{012}$) to give different H fields at the slots while maintaining resonant frequency.
- Designate the AC and CC as 1,2 in the left hand side of Fig. 1. Then the coupling ($k_{12}$) is proportional to the slot size, the product $H_1 * H_2$, divided by $\sqrt{(U_1 U_2)}$. The slot size is calculated by a special routine within the program CCT [2], developed by General Atomics. For the $\pi/2$ mode the energy in cavity 1 is related to the energy in cavity 3 by the following relationship: $k_{12}\sqrt{U_1} = k_{23}\sqrt{U_3}$.
- Slots leading from the end of CC (2) into the center cavity (3) are semi-circular. This is done to enhance the H fields near the slot. See Fig. 4 below. The slot protruding into the waveguide enhances the action of the fields.

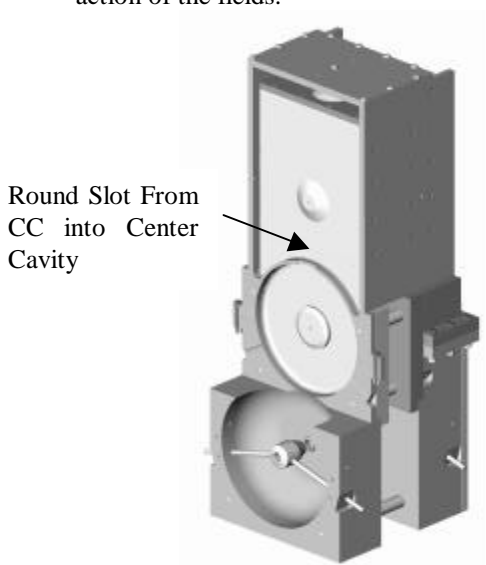

Round Slot From CC into Center Cavity

Figure 4. Cross-section of Model for Test.

- Coupling to an RF feed waveguide is accomplished by means of the slot labeled Center Cavity Power Feed Slot (see Fig. 1). Two slots are incorporated.
- A chamfered post on the CC reduces the fields at the edge of the post. Also by reducing the post diameter, the tuning sensitivity is reduced. See Fig. 5 below.

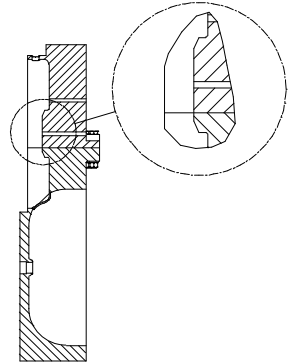

Figure 5. Chamfered CC Posts for Reduced Sensitivity.

## 4 LOCATION OF COUPLING CAVITY

The coupling cavity slot is placed in a high magnetic field area. Elevating the position of the CC relative to the center axis of the AC insures there is only magnetic coupling through the slot. If the coupling slot is located further down from the top of the AC both electric and magnetic coupling terms are present. These terms are of opposite sign and add complexity to the design calculations.

Raising the elevation of the coupling cavities introduced some intricacies to the mechanical design. The CC has an unsymmetrical split that is no longer through the center line (see Fig. 1). If the CC had a low elevation, chamfering the slot would have cut into the second half of the CC.

## 5 GENERAL DESIGN PROCEDURES

The 2D cavities, without slots, are designed so that after the coupling slots are cut, giving the proper coupling, the structure resonates at the proper $\pi/2$ mode frequency. This usually means that a series of iterations are performed to arrive at a self-consistent solution. This procedure is outlined in [2]. In addition for the Coupled Cavity Drift Tube Linac, posts connect the drift tube (see Fig. 4). These posts are used for support and act as cooling passages to the drift tube. Coupling slots lower the frequency, while posts raise the cavity frequency. Similarly this effect is included in the iterative procedure described in [2].

## 6 AC TO CC COUPLING

When apertures (slots) are cut in a wall common to adjacent RF cavities, they will be coupled magnetically,

electrically, or a combination depending upon slot location. The magnitude of the coupling depends on the fields in the two cavities at the location of the slot and on their stored energies. These fields and stored energies are calculated for unperturbed cavities (without slots). If the cavities are axially symmetric, 2-D codes can be used or, if the geometries are simple (such as rectangular), hand calculations are possible. Most of the coupling is magnetic:

$$k_{mag} = \frac{p\mu_0 l^3 e_0^2}{3(K(e_0) - E(e_0))} \frac{H_i H_j \exp(-\alpha_H \cdot t)}{\sqrt{U_i U_j}}$$

where: $\mu_0$ is the permeability of free space.
$\varepsilon_0$ is the permittivity of free space.
$H_i$ ($H_j$) is the magnetic field of the 1$^{st}$ (2$^{nd}$) cavity
$E_i$ ($E_j$) is the electric field of the 1$^{st}$ (2$^{nd}$) cavity.
K and E are elliptical integrals of the 1$^{st}$ kind.
$U_i$ ($U_j$) is the stored energy of the 1$^{st}$ (2$^{nd}$) cavity.
$l$ is the half-length of the slot.
$\alpha_H$ is a damping factor for evanescent modes in elliptical waveguides.
t is the slot depth.

$$e_0 = \sqrt{1 - \left(\frac{w}{l}\right)^2} \text{ see Fig. 6 below.}$$

In Fig. 6 full lengths and widths vs. half lengths and widths are defined respectively with upper and lower case letters. In the coupling equations the length L is always taken in the direction of the magnetic field. In our geometry $l$ is the major axis of the ellipse.

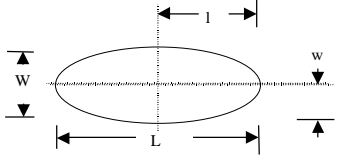

Fig. 6. Definition of Full and Half, Lengths and Widths.

A perturbation can be used to estimate the frequency shift of a single slot in a cavity [3]. The frequency shift is

$$\Delta f_{mag} = f_{noslot} \frac{p\mu_0}{12} \frac{l_1^3 e_0^2}{(K(e_0) - E(e_0))} \frac{H^2 \exp(-\alpha_H \cdot t)}{U}.$$

## 7 CC TO CENTER CAVITY COUPLING

From Reference [4] expressions can be derived to eliminate $H_j$, $U_j$ from the above coupling equations.

$$\frac{\overline{H_{center\,cavity}}}{\sqrt{U_{center\,cavity}}} = \frac{8ba}{w\mu_0 \sqrt{3e_0 a l_g b}} \sin\left(\frac{pL_{end}}{2a}\right).$$

The field $\overline{H_{center\,cavity}}$ is integrated over the slot length $L_{end}$.

## 8 CENTER CAVITY TO WAVEGUIDE IRIS FEED

The waveguide coupling factor ($\beta$) for a cavity coupled to a waveguide can be defined as the ratio of the power emitted from the cavity into the waveguide (through a slot) to the power dissipated in the cavity walls. This calculation again lends itself to perturbation analysis [5].

The waveguide coupling factor is also related to the voltage standing wave ratio (VSWR) in the waveguide. If the waveguide is over-coupled, $\beta$ is equal to the VSWR, while, if the waveguide is under-coupled, $\beta$ is equal to 1/VSWR.

For a rectangular slot, the equivalent area is placed in an ellipse, with the major axis one-half the length of the rectangular slot.

$$\beta = \frac{p^2 Z_0 k_0 \Gamma_{10} e_0^4 l_{wg}^6 \exp(-\alpha_H \cdot t)\left(\overline{H_{wg}}\right)^2}{9ab(K(e_0) - E(e_0))^2 P}$$

where: $Z_0 = \sqrt{\frac{\mu_0}{e_0}}$ $\Omega$, and $k_0 = 2\pi/\lambda$

P = Total power dissipated

The power is related to the total power per feed. One feed supplies power for several AC cells and an associated number of bridge coupler center cells. The power for the AC cells comes from Superfish. The power for the center cavity cells may be calculated by using Ref. [4].

## 9 CONCLUSION

We have developed a bridge coupler employing a 3/2 $\lambda_g$ waveguide-center cavity. The design incorporates a split in the upper half allowing small segments of the accelerator to be removed. The design has the ability to stretch the distance between AC cavities. Using the design code developed by General Atomics gives the designer greater freedom.